\documentclass[12pt]{article}
\begin{document}

\title{{\bf Religious and Scientific Faith in Simplicity}
\thanks{Alberta-Thy-19-08, arXiv:0811.0630, paper written for the
2008 November 21-22 session of the Science and Religion group at New
York University} }

\author{
Don N. Page
\thanks{Internet address:
don@phys.ualberta.ca}
\\
Theoretical Physics Institute\\
Department of Physics, University of Alberta\\
Room 245B1 CEB, 11322 -- 89 Avenue\\
Edmonton, Alberta, Canada T6G 2G7
}
\date{(2008 November 4)}

\maketitle
\large

\begin{abstract}
\baselineskip 24 pt

\hspace{.20in} Both religion and science start with basic assumptions
that cannot be proved but are taken on faith.  Here I note that one
basic assumption that is rather common in both enterprises is the
assumption that in comparing different hypotheses that all equally
explain the observations, the simpler hypotheses are more probable
(Occam's razor or the law of parsimony).  That is, explanations should
be made as simple at possible (though no simpler, since then they would
not explain what is observed).

\end{abstract}
\normalsize

\baselineskip 18.5 pt

\newpage

\section{Introduction}

\hspace{.20in} Religion and science are two different ways of viewing
the world, often seen as contradictory.  Although there are indeed
differences in the ontologies they consider and in the epistemologies
they employ, they are both human methods of cognition that employ
overlapping assumptions and methodologies.  In particular, here I wish
to emphasize how both religion and science share an unproved assumption
in their common element of faith in simplicity, that simpler
explanations for our observations are generally better.  There is also
the related tacit assumption that the world is at least partially
intelligible.

Human beings, and indeed their biological ancestors, have long noticed
that there are apparent regularities in the world.  Babies notice that
parents' faces appear over and over.  Objects placed under a cover
reappear when taken out (``object permanence'').  Objects thrown up into
the air fall down.  The sun is observed to rise and set each day (at
least if one is outside the arctic or antarctic and can see a level
horizon in clear weather).  Lightning is accompanied by thunder.  One
feels warmer standing in the sunshine.  Feelings of hunger are reduced
by eating food.  Sex often leads to reproduction.  (And by reducing
venereal disease and its effects on the unborn, high fidelity leads to
sound reproduction.)  These and a huge host of similar appearances of
regularities lie deep within the human psyche and language.  Logically,
things did not have to be so simple, but in fact many things appear to
have lawful behavior.

The behavior of other humans (and animals) is more complicated than most
of these simpler examples, but humans learned to recognize regularities
there as well.  Mothers will feed crying babies, so babies recognize
that crying can often lead to being fed.  Humans of nearly all ages
recognize that they are more likely to receive something they want from
someone with a smile than from someone with a frown.  Even dogs learn to
come to people offering food and to avoid people speaking with an angry
voice.

Humans recognized that a lot of complex behavior they observed could be 
attributed to willful causal agents, such as other people and animals. 
For humans dealing with lions, it was useful to act as if the lions
wanted to eat the humans.  In the reverse situation, for humans dealing
with prey the humans wanted to eat, it was useful to act as if the prey
did not want to be killed and would escape unless the humans sneaked up
upon them.  For humans seeking companionship with other humans, an
isolated pillar of smoke could be interpreted as the willful action of
other human agents in lighting a campfire.  In less pleasant
circumstances, a man seeing others rushing toward him while brandishing
spears might usefully interpret this as others wishing to kill him.  The
assumption of willful causal agency became highly useful in trying to
make sense of the  world.

Some regularities in the world seem so simple, such as the fact that 
unsupported objects fall, that willful causal agency is probably not the
first thing that comes into mind in trying to explain them.  They are
usually just taken to be facts about the world that need not be
attributed to causal agents.  Other regularities, such as being given
something by another human when asked, are sufficiently complex that
they are often ascribed to willful causal agency.  But there are also
intermediate cases, such as thunderstorms and floods, that are not so
obviously simple as falling objects and yet are neither so obviously the
result of willful causal agency.

Rather than ascribing separate accounts for all these activities, humans
implicitly assumed simplicity in trying to minimize the number of
qualitatively different accounts.  Other than simple activities like
falling objects that might be accepted as basic, humans tended to
characterize more complex activities as all being the result of willful
causal agency.  This explanation is still widely accepted for human
activity, but hundreds of years ago it was also widely accepted for
other complex behavior such as thunderstorms and floods.  These were
attributed to the willful causal agency of various gods.

My claim is that in doing this, humans were not taking a perversely
complex view of reality, ascribing different causes for everything, but
rather they were attempting to simplify their understanding of the world
by ascribing all complex behavior to one single class, willful causal
agency.  So I would think that the move toward a religious viewpoint,
that thunderstorms, floods, etc., were caused by unseen gods, was not a
move away from a search for simplicity in explanations, but rather
a move toward simplicity.

Long ago many religions ascribed different willful causal agencies for
different activities in the world, such as one god for thunderstorms and
another for floods.  However, there was a move toward a simplification
of the picture by reducing the number of gods.  The ultimate
simplification, at  least within the picture of willful causal agencies,
was the development of monotheism and its view of one single God who was
responsible for everything.  (I also do not want to discount the fact
that the Hebrews, who were among the world leaders in developing
monotheism, claimed that this was revealed to them by various religious
experiences and acts in their history.  If one accepts these
experiences, then monotheism may be also seen as the simplest
explanation of these experiences.  But whatever the account, it seems
that it was an underlying faith in simplicity that helped lead to the
acceptance of monotheism.)

Now of course there is a strong movement, promoted by people such as
Richard Dawkins, to reduce the number of gods even further, from one to
zero.  Indeed, much of {\em The God Delusion} \cite{Dawkins} sounds like
a Hebrew Old Testament prophet railing against false gods, except that
Dawkins does not believe in any true God.  It is certainly a matter of
debate whether the simplest explanation of all that we observe is one
God or zero, and I shall discuss this below.  However, here I want to
focus on the apparent fact that the assumption or faith in simplicity
has influenced not only science but also religion.

Indeed, in the West, modern science grew out of not only the Greek
rational heritage (another quest for simplicity) but also the
Judeo-Christian heritage of a single God of law and order
\cite{Hooykaas}.  It was this religious conception of an orderly world
created by one God that led to the development of science.

Although I am certainly not an expert on medieval philosophy and
theology, it seems that during the medieval period the idea of God's
simplicity came to the fore.  A prominent example is the {\em Summa
Theologia} of Thomas Aquinas (c.\ 1225-1274) \cite{Aquinas}, who taught
that God has no composition of parts.

The theological view of the simplicity of God was shortly afterward
applied to the preference for the simplicity of all explanations. 
Although there are predecessors, this preference for simplicity is
commonly attributed to William of Ockham (c.\ 1288-1348), an English
Franciscan friar and scholastic philosopher.  This principle is called
Occam's razor, essentially the same as the `law of parsimony,'' that
``entities must not be multiplied beyond necessity.''  Although it has
always implicitly been a part of human reasoning, it has become an
explicit basis for science.

Thus I would argue that faith in simplicity did not begin with modern
science.  It has been an implicit element of reasoning for nearly all
humans.  It was developed particularly explicitly within religion and
theology, and then it was applied explicitly within the modern science
that grew out of the Judeo-Christian religious tradition.

\section{Simplicity within Bayesian Reasoning}

\baselineskip 18 pt

A goal of science is to develop theories to explain observations.  One
would like to find a theory that is highly probable in view of the
observations, that is, a theory $T_i$ with high posterior probability
$P(T_i|O_j)$, given the observation or set of observations $O_j$ that
the theory is supposed to explain.  This posterior probability is given
by Bayes' theorem as
\begin{equation}
P(T_i|O_j)=\frac{P(T_i\& O_j)}{P(O_j)}
=\frac{p(T_i)P(O_j|T_i)}{\sum_k p(T_k)P(O_j|T_k)}. \nonumber
\end{equation}
where $p(T_k)$ is the prior probability of the theory $T_k$ and
$P(O_j|T_k)$ is the conditional probability of the observation $O_j$
{\it if} the theory $T_k$ is correct.  A sufficiently formulated theory
should, in principle, enable this conditional probability to be
calculated, so logically it is part of any complete theory.

However, the prior probability $p(T_k)$ is not part of the theory $T_k$
itself but is the probability one would assign to the theory in the
absence of any observations such as $O_j$.  It is this probability that
the theory itself cannot provide.  Its assignment is essentially
subjective, in that it is not objectively determined.  However, the
posterior probabilities for the various theories, after the observation
or set of observations, depend  crucially upon these prior
probabilities.  It is effectively here that science asserts its faith in
simplicity by assigning higher prior probabilities to simpler theories. 
Nevertheless, one cannot prove that this should be the case or determine
the prior probabilities from the observations.

To consider a simple example, consider a finite sequence of integers as
an observation, and consider theories that give the rules for
constructing an infinite sequence that starts with this finite
sequence.  For example, think of the finite sequence 1, 2, 3, 4, 5, 6,
7, 8, 9, 10.  If this were to be extended to an infinite sequence, what
would you say that the theory or rule is, and what numbers does it
predict for the next three entries in the sequence?  Perhaps most people
would say something to the effect that this finite sequence is the
beginning of the sequence of natural numbers, otherwise known as the
whole numbers, the counting numbers, or the positive integers, with each
new element in the sequence being one more than the previous one.  Then
the next three elements would be 11, 12, 13.

However, the finite sequence given above is also the beginning of the
Niven (or Harshad) numbers \cite{Niven-sequence}: numbers that are
divisible by the sum of their digits.  Then the next three entries would
be 12, 18, 20.  Alternatively, this finite sequence is the beginning of
highly composite numbers \cite{composite-sequence}: numbers whose prime
divisors are all less than or equal to 7.  This sequence would give the
next three entries as being 12, 14, 15.  Therefore, just from the
observation of the first ten entries of an infinite series, one cannot
definitely conclude what the entire sequence is or even what the next
few entries are.

Nevertheless, probably most people would say that the sequence is the
natural numbers (or words to that effect), and that the next three
entries are 11, 12, 13.  Why is that?  I would argue that it is because
the infinite sequence of natural numbers seems simpler than either the
Niven numbers or the highly composite numbers.  Therefore, based upon
the unproved faith in simplicity, one favors the idea (assigns higher
prior probability to the  idea) that the infinite sequence is the
natural numbers rather than the Niven numbers or the highly composite
numbers.

Faith in simplicity is the basis for scientific induction, generalizing
from a finite number of examples to a wider regularity.  We see the sun
rise for many days and form the induction that it will continue to rise
day after day.  We note that objects thrown into the air fall down and
conclude that such objects will always fall.  These inductions are
certainly not logical deductions that must inevitably follow from the
observations, but  predictions based on a belief in simplicity.  Of
course, we may find that our predictions are wrong, as when we go north
of the Arctic Circle at the beginning of winter, or when we use a rocket
to escape the earth's gravitational field.  But then we try to look for
the simplest possible reasons why our original predictions failed and to
look for more general simple predictions that apply even more generally.

One application of our faith in simplicity is our belief that there was
an actual existent past.  In some sense, all that one has direct
epistemological access to is one's own conscious perception at the
present.  We effectively use our faith in simplicity (usually
unconsciously; apparently we are just wired to think that way) to assume
that the experience of an apparent memory of a past experience denotes
the actual existence of that experience in the past, though each of us
really has direct access only to his or her present conscious
experience.

On a recent flight from Vancouver to Seoul, I flew nearly directly over
an abandoned village in Alaska (Holikachuk) where I had lived from the
ages of one to nine but had left just over 50 years ago and had not seen
since.  I was impressed with the strong correlation between my memories
of the layout of the Innoko River and connecting sloughs at the village
with the present appearance (now as I write this, with my memory of the
appearance from the airplane).  I have no proof that this correlation
within my present perception of what I interpret as memories from over
50 years ago and of what I interpret as memories from my recent airplane
flight is not just some random correlation, but it is much simpler to
suppose that I actually did have experiences of a river and sloughs 50
years ago and that these geological features have persisted and given me
correlated visual impressions 50 years later.  (Of course, the
correlations are not perfect.  I also spotted from the airplane the
location of the bend of the river, shortly below the mouth of the
Iditarod River whose name has become much more famous than it was when I
lived in Holikachuk, where I first learned the distinction between left
and right, but now that bend has been cut off by an oxbow, or at least
that is my simple interpretation.  Fortunately, my concept of left and
right has been transferred to other bases and no longer relies on that
bend in the river that has disappeared.)

Besides the extrapolation from present conscious experience to the
assumption of the existence of the past, there is the extrapolation to
the assumption that there is an external world causing much of one's
conscious experience.  For example, I believe the correlation between
the part of my present perception that I interpret as memories of over
50 years ago in Holikachuk and the part that I interpret as memories of
my recent airplane flight over Alaska is not just caused by features of
my mind but is also partially caused by features in the external world,
such as the general persistence of the layout of the Innoko River (other
than the bend that got cut off) and the sloughs near the location of
Holikachuk, and the fact that light propagates in nearly straight lines
from such features and is focused by my eyes.

In addition, there is also the extrapolation from the sense that one is
a willful causal agent to the belief that other willful causal agents
also exist (e.g., other people), which I have mentioned above. 
Philosophers have noted that there is no logical proof in the existence
of other minds (just as there is no proof in the existence of the past
or even of an external physical world).  However, to me other minds seem
to be a consequence of the simplest explanation of much of my present
conscious experience (though I personally might doubt that the simplest
explanation of the existence of other minds, that is, of conscious
experiences of others, implies that such minds are really willful causal
agents in the sense of being the ultimate causes of anything, but that
will be discussed later).

These examples are of beliefs that presumably most humans have held from
pre-historic times, so that they are so deeply ingrained in our psyches
that it is hard to consider alternatives.  However, there are more
recent examples of the influence of our faith of simplicity in the area
of science where it is easier to conceptualize alternative beliefs.

One example from modern science would be the formulation of Newton's law
of universal gravitation, generalizing the observation of falling
objects on the earth, first to the moon that falls toward the earth to
orbit it rather than going off in a straight line.  Then Newtonian
gravity was applied to the orbits of planets around the sun, giving a
simpler explanation of Kepler's laws that up to then had been the
simplest description of the orbits.  Furthermore, Newtonian gravity was
applied to the gravitational forces between planets to explain even the
small deviations from Kepler's laws that had been observed.  A
brilliant culmination of its power was its use to explain the motion of
Uranus by the postulation of a new planet which was then discovered and
named Neptune.

In the late nineteenth century it was found that even after including
the gravitational attractions on Mercury from the other planets as well
as from the sun (which make up the largest part of the discrepancy of
its orbit from that given by Kepler's laws), its orbit did not quite fit
the Newtonian prediction.  Explaining this deviation correctly became a
major triumph of Einstein's theory of general relativity, though that
theory of gravity was formulated basically to reconcile gravitational
theory with special relativity instead of specifically to solve the
problem of the orbit of Mercury.  Einstein's theory applied more
generally than Newton's theory of gravity.  Although from the viewpoint
of most undergraduate mathematics it looks more complicated than
Newton's theory, from a more advanced mathematical viewpoint that
encompasses time as well as space in the unified structure of spacetime,
Einstein's theory appears conceptually even simpler than Newton's.

Now we know that even Einstein's general relativity theory is not the
last word on gravity, since it does not incorporate quantum theory,
which has been very strongly confirmed for nongravitational interactions
and most simply should apply to gravitation as well.  It is not yet
understood what a complete theory of quantum gravity is, though the
partially formulated ideas of string theory at present appear to be the
best candidate for such a theory (perhaps also with contributions from
other approaches such as loop quantum gravity).  The mathematics of what
we presently know of string theory looks very forbidding, so I am not
sure that there is any person on earth (well, possibly Edward Witten is
an exception) for whom string theory looks simpler than general
relativity, but I think a hope of many physicists is that with
sufficiently sophisticated mathematics, string theory might ultimately
be seen as simpler than general relativity (or at least simpler than
general relativity plus quantum theory, which at present we do not know
how to reconcile).

These examples can also be used to illustrate that the preference for 
simplicity should only be applied to different theories that explain the
same set of observations, and ultimately we would like theories that
explain as much as possible.  If one only looks at how an object falls
at one location on earth, it is simplest to say that it experiences a
downward gravitational force equal to its mass multiplied by the local
acceleration of gravity, $F=mg$.  If one looks at a large (non-quantum)
objects moving much slower than the speed of light at an arbitrary
location in a general gravitational field that is too weak to cause
anything to move close to the speed of light, one cannot just use $F=mg$
with a constant $g$, and then it seems simplest to use Newton's law of
gravitation.  If one sticks to non-quantum objects and sources of the
gravitational field but allows speeds close to the speed of light and/or
gravitational fields strong enough to cause objects to move at such high
speeds, then Newton's theory is not accurate, and the simplest theory
that appears to be accurate is Einstein's theory of general relativity. 
If one goes further to quantum sources of the gravitational field, then
general relativity is also not applicable, and then the simplest theory
is not yet known but might be some completion of something like string
theory.

At present, since we do not fully understand string theory or any other
candidate theory of combining quantum theory and all the known
interactions (including gravity), we use a hodge-podge of ideas, such as
general relativity for gravity and quantum field theory for other known
interactions (e.g., electromagnetism, weak interactions, and strong
interactions).  Besides the fact that we do not know how to apply the
hodge-podge consistently to all conceivable situations (particularly to
the beginning of the universe and to the complete Hawking evaporation of
black holes), we recognize that the hodge-podge is conceptually not very
simple, so we desire a simpler unifying theory.

Another point to be made is that the simplicity desired is in the basic
principles of the fundamental formulation, not in the simplicity of
being able to calculate the consequences easily.  In astronomy, the
Ptolemaic system was a remarkably brilliant method of calculating most
of the motion of the planets to as much accuracy as was generally needed
to explain the naked-eye observations, at least before Tycho Brahe's
careful observations that were the most precise before the invention of
the telescope.  As I understand it, the calculations were actually
harder to make using Kepler's laws, but even without the improvement
that it gave to explain some of Tycho's crucial observations, it was
conceptually much simpler than the Ptolemaic system.  (That system
turned out with hindsight to be a very good approximation to Kepler's
laws, but when it was formulated, the conceptually simpler Keplerian
laws were not known, so the approximation to them was cleverly
discovered before the conceptually simpler generalization.)  Even before
Kepler found his laws that fit the new data of Tycho Brahe better than
the Ptolemaic system, some, but not all, astronomers had turned to the
Copernican system for its greater conceptual simplicity over the
Ptolemaic system, even though with its originally circular orbits it
did not actually fit the data nearly so well as the Ptolemaic system.

Somewhat analogously, when one went from Kepler's laws to Newton's, one 
gained the conceptual simplicity of being able to derive all three of 
Kepler's laws from one inverse-square acceleration of each planet toward
the sun, but when one included the further consequence that there are
also forces between the planets, the calculations became much harder
than that needed with Kepler's laws (themselves harder for manual
calculations than the calculations in the Ptolemaic system).  Indeed,
for centuries it was not known whether an idealized Solar System of
point masses for the sun and  planets is stable or not in Newton's
theory.  Newton thought that it wasn't stable and invoked God to set
things aright, and Laplace erroneously thought he had proved it was
stable and said he had no need for the hypothesis of God.  In the 20th
century numerical calculations strongly suggest, without being quite a
proof, that the Solar System is not stable but is just old \cite{LFHM},
so that the objects with shorter timescales for their instabilities have
been ejected, leaving objects that are relatively stable for timescales
longer than the finite age of the Solar system.

Similarly, when one went from Newton's laws to Einstein's, one gained a
conceptual simplicity of being able to derive gravitation from the
curvature of spacetime, rather than by postulating forces acting at a
distance, but the calculations became even more difficult.  People are
just beginning to gain enough computer power and skills to calculate how
much energy is radiated in gravitational waves as two black holes spiral
together and coalesce.  When one makes the further move from Einstein's
general relativity to string theory, no one is even close to being able
to make the calculations in most situations, such as precisely what
happens in the final stages of the Hawking evaporation of a black hole.

Now there is growing evidence that string theory may lead to a huge
landscape of perhaps $10^{100}$--$10^{10\,000}$ different `vacua,'
different effective low-energy laws of physics
\cite{Susskind,Vilenkin,Carr}.  These different vacua are analogous to
the DNA for different living organisms.  They set limits but do not
determine what happens, which also requires the initial conditions (or
the quantum state of that vacuum), just as an organism is not completely
determined by its DNA but also has environmental influences.  The
conjecture is that the full quantum state of the entire universe would
not be concentrated on a single vacuum but would be spread over most or
all of this landscape, populating it with a multiverse of actual vacua
or subuniverses.  If this is indeed the correct picture, each individual
vacuum might be rather complex (e.g., requiring at least 100--10\,000
digits to specify), but the entire string theory and quantum state that
determines the properties of the entire multiverse might be much
simpler.

An analogy to such a multiverse might be the set of all natural numbers
(the counting numbers or positive integers), which is in its entirety
conceptually a very simple set, but nearly all of its individual members
are  very large and very complex integers.  For example, if any long and
complex  book (or even a whole library of books) is encoded into binary
digits and then this sequence of binary digits is re-interpreted as a
binary integer, one would have encoded the book or library into this
single integer, which is much more complex than the set of all
integers.  These examples show that it is easy for the whole to be
simpler than the parts.  (When the whole is  very simple, as it is for
the set of all natural numbers and as it might be for the entire quantum
state of the universe, then the complexity of each part would largely
lie in the complexity of the specification of the choice of that part
out of the whole.)

A theistic analogue of the putative simplicity of the entire universe
would be the medieval concept that God is a simple being who is
omniscient and omnipotent.  Almost all of the individual things God
knows and can do may be extremely complex, but the knowledge of all
truth, and the ability to do anything not logically inconsistent, may be
considered to be quite simple.  I believe the current quest for ultimate
simplicity in our scientific theories is highly analogous to the
medieval desire for ultimate simplicity in the concept of God.

\section{The Complexity and Probability of God}

\baselineskip 20 pt

\hspace{.20in} Richard Dawkins, in {\em The God Delusion}
\cite{Dawkins}, argues against the existence of God by saying that God
would have to be extremely complex.  Since his arguments are not very
tightly stated, I formulated the heart of Dawkins' argument as a
syllogism and then, with the help of an email exchange with several
colleagues, especially William Lane Craig, I revised it to the following
form:

\begin{enumerate}
\item A more complex world is less probable than a simpler world.
\item A world with God is more complex than a world without God.
\item Therefore a world with God is less probable than a world without
God.
\end{enumerate}

After circulating this form, I did get the obviously hurried reply from
Dawkins:  ``Your three steps seem to me to be valid.  Richard Dawlkins
[sic]'' (1 February 2007).

Now that I have summarized Dawkins' basic argument in a brief form that
he seems to agree with, modulo typos, one can ask whether Dawkins is
right.  The conclusion of the syllogism seems to follow from the two
premises (or at least I have intended this to be the case), so it is a
question of whether the premises are correct.

One might question whether complexity is improbable, an unproved
assumption.  There is also the fact that complexity depends on
background knowledge and may be only subjective.  For example, David
Deutsch \cite{Deutsch1} has emphasized to me that ``complexity cannot
possibly have a meaning independent of the laws of physics. \ldots If
God is the author of the laws of physics (or of an overarching system
under which many sets of laws of physics are instantiated---it doesn't
matter) then it is exclusively God's decision how complex anything is,
including himself.  There neither the idea that the world is `more
complex' if it includes God, nor the idea that God might be the
`simplest' omnipotent being makes sense.''  

This argument makes sense to me, but it did have the effect of shaking
my fundamentalist physicist faith in the simplicity of the laws of
nature.  However, more recently Deutsch has pointed out \cite{Deutsch2}
that these considerations do not mean that the concept of the simplicity
of the laws of physics is circular:  ``I don't think it's circular,
because the fact that simplicity is determined by the laws of physics
does not mean that all possible laws are `simple' in their own terms.'' 
So I suppose one might still ask whether in a universe apparently
governed by simple laws of physics, God appears to be simple.  However,
since as Deutsch notes, God could have made Himself appear to have
arbitrary complexity, it is a bit dubious to say that His probability is
determined by His complexity.

Nevertheless, since we scientists (and indeed most others) prefer
hypotheses that are ultimately simple, we might for the sake of argument
grant the first premise I have ascribed to Dawkins and ask whether the
second premise is correct.  Again Deutsch's comments should lead us to
be cautious in drawing such conclusions.

However, even if we take the na\"{\i}ve view that one can define the
complexity of God (say with respect to the laws of physics in our
universe), then it is still not obvious that God is complex, or that He
would add complexity to the world.  Perhaps God is indeed simple
\cite{Swinburne}.

If God were necessary, then He would have no complexity at all, because
no information would be needed to specify Him.  Even if God were
contingent, He might be simple.  For example, perhaps God is the best
possible being (assuming sufficient background knowledge that this
apparently simple definition uniquely specifies some possible entity,
though it is certainly unclear that our background knowledge within this
universe is sufficient for this).  Even if it is not necessary for such
a God to exist, He might be simple (if simplicity can indeed be
defined).

Even if one concedes that the philosophical idea of God might be simple,
there is the question of whether God is simple in traditional
monotheism.  At first sight, the God of the Bible and of the Koran seems
complex.  But analogously, Earth's biosphere seems complex.  However, the
full set of biospheres arising by evolution in a huge universe or
multiverse with simple laws of physics might be simple.  Similarly, the
limited aspects we experience of God might be complex, but the entirety
of God might be simple.

I mentioned near the beginning of this paper that one reason that humans
developed the conception of gods or God was to try to get a simple
explanation of their observations and experiences in terms of willful
causal agents.  Now with the growth of a rather mechanistic view of the
universe in  science, reasons have arisen to doubt the existence of
willful causal agents within the universe.  For example, if quantum
theory gives a complete  description of the physical universe and does
not have any collapses of the wavefunction that would violate the
quantum laws of unitary evolution, there does not seem to be much room
for incompatibilist free will within the universe (though I suppose one
could postulate that free will agents could help choose the quantum
state and its evolution).  Also, in physics causality applies equally
forward and backward in time, so there seems to be no fundamental
distinction between cause and effect.  Thus, one might doubt the usual
assumption of a unidirectional causation in which it is clear what the
causes are and what are the effects.

These two reasons for doubting the fundamental existence of willful
causal agents within the universe might be interpreted as attacks on the
scaffolding humans have used to extrapolate to the concept of God as the
ultimate willful causal agent.  Indeed such arguments do lead me to
question arguments from willful causation to the existence of God. 
However, it has also occurred to me that the concept of God as the
ultimate willful causal agent may be correct even if the scaffolding we
have used to arrive at this concept is incorrect.  What if God as the
personal Creator of the universe created us in His image, not as truly
having free will in the incompatibilist sense and being truly causal
agents, but as beings whose thoughts and actions mirror God's true free
will and causation?  God might have created us to experience some of the
feelings that He has as a willful causal agent, even though for us those
feelings would be somewhat illusory, in that we ourselves do not really
have free will (at least in the incompatibilist  sense) and are not
really the cause of anything, since (in what I regard as the simpler
view) it is only God that is the true cause of everything.

\section{Summary}

In conclusion, I believe both religion and science share a common
underlying unproved faith in simplicity.  One might argue that the
assumption of simplicity has proved itself by working in the past.  But
this argument depends on the assumption that what has worked in the past
is true and will work in the future, which is essentially a special case
of the assumption of simplicity that I am arguing is not proved.  Of
course, one can take the experience of its working in the past as
evidence for its truth, just as one can take religious experience as
evidence for the truth of religious beliefs, but ultimately one cannot
prove this assumption and can only accept it by faith.

Without something like this faith in simplicity, one can hardly claim to
know anything, so I would say that our epistemology implicitly depends
upon this unproved faith.  Although it is not quite the same thing, this
faith in simplicity as the basis for knowledge seems somewhat analogous
to what is written in Proverbs (1:7, 9:10, and 111:10) about faith in
God and moral understanding:  ``The fear of the Lord is the beginning of
wisdom.''

\section*{Acknowledgments}

\hspace{.20in}	I am indebted to discussions with Andreas Albrecht,
Denis Alexander, Stephen Barr, John Barrow, Nick Bostrom, Raphael
Bousso, Andrew Briggs, Jeffrey Brower, Peter Bussey, Bernard Carr, Sean
Carroll, Brandon Carter, Kelly James Clark, Gerald Cleaver, Francis
Collins, Robin Collins, Gary Colwell, William Lane Craig, Paul Davies,
Richard Dawkins, William Dembski, David Deutsch, the late Bryce DeWitt,
Michael Douglas, George Ellis, Debra Fisher, Charles Don Geilker, Gary
Gibbons, J.~Richard Gott, Thomas Greenlee, Alan Guth, James Hartle,
Stephen Hawking, Rodney Holder, Chris Isham, Werner Israel, Renata
Kallosh, Klaas Kraay, Karel Kucha\v{r}, Denis Lamoureux, John Leslie,
Andrei Linde, Robert Mann, Don Marolf, Alister McGrath, Ernan McMullin,
Tom Nagel, Gerard Nienhuis, Andrew Page, Cathy Page, John Page, Gary
Patterson, Alvin Plantinga, Chris Polachic, John Polkinghorne, Martin
Rees, Hugh Ross, Henry F.~Schaefer III, Paul Shellard, James Sinclair,
Lee Smolin, Mark Srednicki, Mel Stewart, Jonathan Strand, Leonard
Susskind, Richard Swinburne, Max Tegmark, Donald Turner, Neil Turok,
Bill Unruh, Alex Vilenkin, Steven Weinberg, Robert White, and others
whom I don't recall right now, on various aspects of this general
subject, though the opinions expressed herein are my own.  My own
scientific research is supported in part by the Natural Sciences and
Research Council of Canada.


\newpage
\begin{thebibliography}{99}

\baselineskip 20 pt

\bibitem{Dawkins} Richard Dawkins, {\em The God Delusion} (Houghton
Mifflin, Boston, 2006).

\bibitem{Hooykaas} Reijer Hooykaas, {\em Religion and the Rise of Modern
Science} (Regent College Publishing, Vancouver, 2000).

\bibitem{Aquinas} Thomas Aquinas, {\em Summa Theologia} (Cambridge
University Press, Cambridge, 1990).

\bibitem{Niven-sequence} Robert G.~Wilson, Sequence A005349 in N. J. A.
Sloane, ed., {\em The On-Line Encyclopedia of Integer Sequences} (2008),
published electronically at
http://www.research.att.com/$\sim$njas/sequences/A005349.

\bibitem{composite-sequence} James A.~Sellers and Michel Lecomte,
Sequence A002473 in N. J. A. Sloane, ed., {\em The On-Line Encyclopedia
of Integer Sequences} (2008), published electronically at
http://www.research.att.com/$\sim$njas/sequences/A002473.

\bibitem{LFHM} Myron Lecar, Fred A.~Franklin, Matthew J.~Holman, and
Norman W.~Murray, ``Chaos in the Solar System,'' Annual Review of
Astronomy and Astrophysics {\bf 39}, 581-631 (2001).

\bibitem{Susskind} Leonard Susskind, {\em The Cosmic Landscape:  String
Theory and the Illusion of Intelligent Design} (Little, Brown and
Company, New York, 2006).

\bibitem{Vilenkin} Alex Vilenkin, {\em Many Worlds in One: The Search
for Other Universes} (Hill and Wang, New York, 2006).

\bibitem{Carr} Bernard Carr, editor, {\em Universe or
Multiverse?} (Cambridge University Press, Cambridge, 2007).

\bibitem{Deutsch1} David Deutsch (private communication, Jan.~22, 2007).

\bibitem{Deutsch2} David Deutsch (private communication, Sept.~29, 2007).

\bibitem{Swinburne} Richard Swinburne, {\em The Existence of God}
(Clarendon, Oxford, 1991).

\end{thebibliography}
\end{document}